%% file: main.tex
\documentclass[conference]{IEEEtran}
\IEEEoverridecommandlockouts
\usepackage{cite}
\usepackage{soul}
\usepackage{amsmath,amssymb,amsfonts}
\usepackage{algorithmic}
\usepackage{graphicx}
\usepackage{textcomp}
\usepackage{xcolor}
\usepackage{subfig}
\usepackage[hyphens]{url}
\usepackage[bookmarks=true,breaklinks=true,letterpaper=true,colorlinks,citecolor=blue,linkcolor=blue,urlcolor=blue]{hyperref}
\def\BibTeX{{\rm B\kern-.05em{\sc i\kern-.025em b}\kern-.08em
    T\kern-.1667em\lower.7ex\hbox{E}\kern-.125emX}}

\newcommand{\REM}[1]{}

\begin{document}

\title{Monad: Towards Cost-effective Specialization for Chiplet-based Spatial Accelerators}
\author{\IEEEauthorblockN{Xiaochen Hao\textsuperscript{1},
Zijian Ding\textsuperscript{1},
Jieming Yin\textsuperscript{2},
Yuan Wang\textsuperscript{1},
Yun Liang\textsuperscript{1,3*}}
\IEEEauthorblockA{\textsuperscript{1}Peking University, \textsuperscript{2}Nanjing University of Posts and Telecommunications}
\IEEEauthorblockA{\textsuperscript{3}Beijing Advanced Innovation Center for Integrated Circuits}
\IEEEauthorblockA{xiaochen.hao@stu.pku.edu.cn, dzj325@gmail.com, jieming.yin@njupt.edu.cn, \{wangyuan, ericlyun\}@pku.edu.cn}
}
\maketitle
\renewcommand{\thefootnote}{\fnsymbol{footnote}}
\footnotetext[1]{Corresponding author.}

\begin{abstract}
Advanced packaging offers a new design paradigm in the post-Moore era, where many small chiplets can be assembled into a large system. Based on heterogeneous integration, a chiplet-based accelerator can be highly specialized for a specific workload, demonstrating extreme efficiency and cost reduction. To fully leverage this potential, it is critical to explore both the architectural design space for individual chiplets and different integration options to assemble these chiplets, which have yet to be fully exploited by existing proposals.

This paper proposes Monad, a cost-aware specialization approach for chiplet-based spatial accelerators that explores the tradeoffs between PPA and fabrication costs. To evaluate a specialized system, we introduce a modeling framework considering the non-uniformity in dataflow, pipelining, and communications when executing multiple tensor workloads on different chiplets. We propose to combine the architecture and integration design space by uniformly encoding the design aspects for both spaces and exploring them with a systematic ML-based approach. The experiments demonstrate that Monad can achieve an average of 16\% and 30\% EDP reduction compared with the state-of-the-art chiplet-based accelerators, Simba and NN-Baton, respectively. 

\end{abstract}

\input{introduction}
\input{background}
\input{model}
\input{optimization}

\input{experiments}
\input{related-work}

\bibliographystyle{IEEEtranS}
\bibliography{references}

\end{document}

%% file: introduction.tex
\section{Introduction}

The slowing of Moore's Law has motivated the industry to embrace chiplet-based designs, where a large monolithic die is broken into multiple smaller dies called ``chiplets''. Small chiplets benefit from high fabrication yields and short development cycles. By leveraging advanced packaging technologies, multiple chiplets can be integrated into the same package, delivering a flexible, high-performance processor at a reasonable cost~\cite{shan2022architecture}.

Meanwhile, as the computing demand for machine learning keeps increasing, spatial accelerators that feature an array of processing elements (PEs) emerge as an efficient platform and also become larger and larger. For example, Cerebras~\cite{cerebras} uses a wafer-scale accelerator to offer cluster-scale throughput. To take full advantage of the performance, the large accelerators are designed to execute a graph of workloads in parallel, e.g., several DNN layers rather than a single convolution~\cite{cai2022deepburning,wei2018tgpa}. Besides, these accelerators are generally organized in a tiled architecture, where multiple identical small cores are interconnected with a high-bandwidth network-on-chip (NoC). Due to their high complexity, these accelerators may experience long development cycles, increased costs, and difficulty adapting to specific applications~\cite{atomic,tangram,zheng2022amos}.

Compared to the large monolithic designs, a chiplet-based accelerator can offer cost reduction and efficiency benefits by using small dies and allowing \textit{packaging-level specialization}. In a tiled architecture, the compute and communication resources are uniformly distributed across the hardware, which may cause resource under-utilization when running a specific workload. The chiplet approach, however, allows designers to integrate specialized small dies into a system to accelerate a graph of workloads. Each chiplet can be designed specifically for one workload with an affordable cost, and the in-package network for interconnecting multiple dies can be specialized to match the communication pattern and bandwidth requirement of the target workload graph. In addition, the chiplet and network designs can be reused across different accelerators~\cite{wang-noc}, amortizing their development and fabrication costs.

Specializing a chiplet-based accelerator poses challenges in architecting the chiplets and assembling them into a package. From an architectural perspective, a \textit{resource assignment} decides the number of processing elements (PEs) and the amount of on-chip buffers in every chiplet. A \textit{dataflow} describes how a workload is parallelized and how its data are moved. These two aspects are critical to acceleration efficiency~\cite{maestro,tenet}. From an integration perspective, the in-package wires cannot provide the same bandwidth and/or energy per bit as on-chip wires, so that an inefficient \textit{in-package network} may become a performance bottleneck. Besides, \textit{packaging technology} plays a pivotal role since it determines not only available bandwidth but also the  fabrication cost. Depending on the performance and cost budget requirement, both traditional (but cheap) and advanced (but expensive) packaging should be evaluated.



We need a comprehensive chiplet-based spatial accelerator design methodology to tap their potential. The existing works revolve around either the architectural or the integration aspects and follow a separate design flow: given chiplets then design packaging~\cite{yin-noc,kite}, or given packaging then design chiplets \cite{simba, nnbaton, sprint, krishnan2022big}. In fact, there is a richer design space if tradeoffs can be made between the architecture and integration design. For example, varying resource assignment affects communication demands and subsequently the choice of packaging. A separate flow would consider one of them at a time, leading to a suboptimal design.

In this paper, we argue that we should consider PP\textbf{P}A as the design objectives, which stands for \underline{p}ower, \underline{p}erformance, \textbf{\underline{p}rice}, and \underline{a}rea. Using cost (price) as a metric can be helpful in directing optimizations, as chiplet approach can reduce cost but at the expense of the others. Partitioning a monolithic chip reduces the cost, but requires adding die-to-die interfaces and thus uses extra energy and area. More advanced (and costly) packaging technology enhances connectivity, but may reduce the budget on chiplets for computing efficiency.

We propose a modeling framework to assess the efficiency of specialized chiplets and their interconnection. The chiplets vary in resources and dataflow, making it essential to model the parallel hierarchy, computing, data access, and data reuse behaviors. We target expediting a graph of tensor workloads by assigning each chiplet a single workload and coordinating multiple chiplets in a pipelined manner. It is critical to model the pipeline efficiency of compute and data transfer stages. In addition, the data transfer efficiency may vary with communication flows, network structures, and available bandwidth in a packaging. Our framework thus models the routing and flow control mechanism of a network to capture this variation. The modeling outputs a performance estimation, thereby enabling the exploration of the design space.

We further propose to co-design the architecture and integration of a chiplet-based accelerator. Specifically, the architecture involves resource assignment and dataflow, while the integration involves network and packaging. Our optimization framework encodes parameters from the above design aspects and automate the exploration with a Bayesian engine.

In summary, this paper makes the following contributions:
\begin{itemize}
    \item We propose a cost-aware design approach to make comprehensive tradeoffs for a chiplet-based accelerator.
    \item We propose a modeling framework to evaluate a chiplet system with specialized architecture and interconnects. 
    \item We develop an ML-based co-optimization framework to couple the architecture and integration design space.
\end{itemize}

Experiments present an average of 16\% and 30\% energy-delay-product (EDP) reduction compared with the state-of-the-art, Simba~\cite{simba} and NN-Baton~\cite{nnbaton}, respectively. We achieve 24\% less latency or 16\% less energy compared with the best of separate architecture or integration optimization.

%% file: background.tex
\section{Background}

\subsection{Hardware and Application Model}
\begin{figure}[t]
    \centering
    \includegraphics[width=\columnwidth]{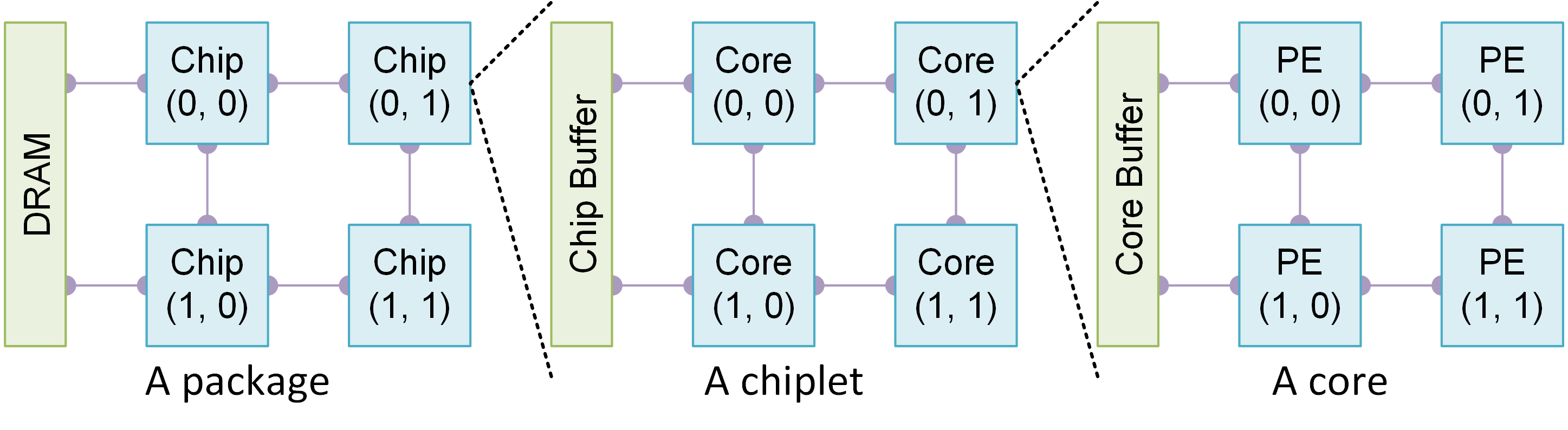}
    \captionsetup{font={small}}
    \caption{A chiplet-based spatial accelerator with distributed compute and memory elements}
    \label{fig:hardware-model}
    \vspace{-0.1in}
\end{figure}

We model a chiplet-based spatial accelerator in three hierarchical levels: a package, chiplets, and cores. As shown in Fig.~\ref{fig:hardware-model}, a package consists of  chiplets, and a chiplet consists of multiple cores. A core has a PE array, in which every PE has a MAC unit. The boundary chiplets are connected with DRAM, and other chiplets load data through the in-package interconnection. The chiplet buffer is shared by all its cores, and the core buffer is shared by all its PEs. Each chiplet can be specialized, i.e., designed for a specific workload, in terms of resources and dataflow. PEs, cores, and chiplets are interconnected. An in-package network (connecting chiplets) can be specialized to match the communication flow of multiple chiplets running different workloads.

We target tensor workloads such as matrix multiplication, convolution, MTTKRP, etc., which can be described using a loop nest, i.e., an operation on tensors. We refer to \textit{dataflow} as the way to orchestrate data movement and data reuse. The dataflow can direct where (a PE in the hierarchy) and when (a sequence) to execute a loop instance and access its associated data. The same data can be reused across interconnected PEs, or across time within the same PE. Modeling dataflow offers an accurate estimation of an accelerator's utilization, latency, and energy consumption~\cite{tenet,maestro}, which facilitates spatial hardware synthesis~\cite{tensorlib,rubick,ems}.

\subsection{Advanced Packaging}

\begin{figure}[t]
    \centering
    \subfloat[Organic Substrate]{
        \includegraphics[width=0.45\columnwidth]{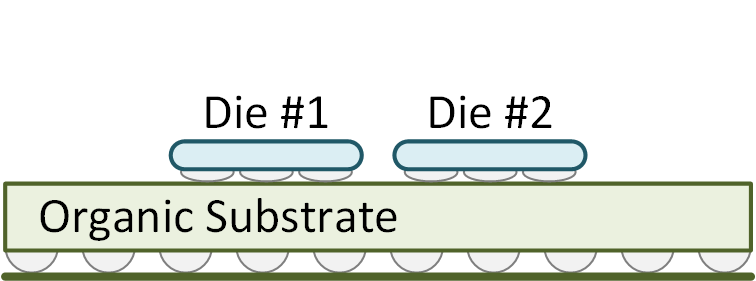}
        \label{fig:organic-substrate}
    }
    \subfloat[Silicon Interposer]{
        \includegraphics[width=0.45\columnwidth]{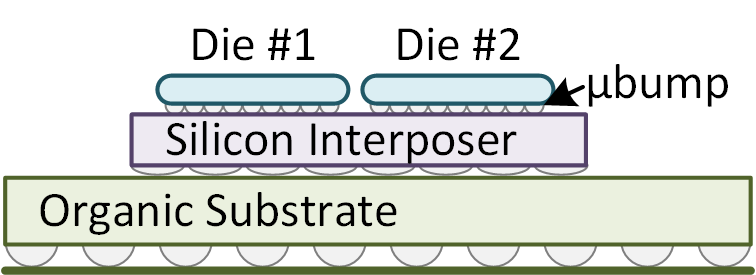}
        \label{fig:silicon-interposer}
    }
    \captionsetup{font={small}}
    \caption{Typical advanced packaging technologies}
    \label{fig:packaging-technology}
    \vspace{-0.1in}
\end{figure}

Advanced packaging technologies, including organic substrate and passive/active silicon interposer, allow the assembling of separately manufactured dies in a package. Multiple dies can be mounted on an organic substrate in Fig.~\ref{fig:organic-substrate}. A silicon interposer, as given in Fig.~\ref{fig:silicon-interposer}, is a die for connectivity on top of which dies are mounted. It interconnects dies with microbumps, which provides a high interconnect density (6$\times$ compared with organic substrate~\cite{ucie}). The active interposer can incorporate active devices, like network routers, to save bump resources for higher connectivity~\cite{yin-noc,wang-noc}. However, a passive interposer is fabricated with costly processes, while an active interposer even relies on standard CMOS processes, causing much-increased costs~\cite{stow-interposer}.

We calculate the total fabrication cost of chiplet-based accelerators by considering costs of die fabrication, die bonding, substrate and interposer fabrication, and additional processes:
\begin{equation}
    \label{equ:cost}
    C_{total} = \sum_{i=1}^N(\frac{C_{die}^i}{y_{die}^i}+C_{bond}) + C_{sub} + \frac{C_{int}}{y_{int}} + C_{proc}
\end{equation}

We consider $N$ dies to be assembled in a package. Both die cost $C_{die}^{i}$ and die yield $y_{die}^{i}$ depend on the die $i$'s area and technology node. The yield is estimated with a negative binomial model. The substrate cost $C_{sub}$ and interposer cost $C_{int}$ is proportional to its area, and the interposer yield $y_{int}$ is estimated similarly to that of a silicon die. In this work, we focus on standard packaging technologies and fabrication cost with an industrial cost model~\cite{icknowledge}, but our method can be extended to industrial variants and consider the non-recurring engineering (NRE) cost with an advanced model~\cite{feng2022chiplet}. 

We demonstrate the impact of chiplet scale and packaging technology on the total fabrication cost by using two typical DNN accelerators, TPU~\cite{tpu} and Gemmini~\cite{gemmini}. TPU uses a 256$\times$256 systolic array within 331$mm^2$ area in 28nm node. Gemmini is assumed to have a 16$\times$16 systolic array within 1.1$mm^2$ area in 22nm. Each TPU or Gemmini chip is used as one chiplet in a system, and we integrate three chiplets with an organic substrate or a passive/active interposer. The total fabrication cost is estimated with Equation~\ref{equ:cost}. We assume the baseline as a large monolithic die offering the same capability as a three-chiplet system, leading to three times the area of a single chip. We normalize its die cost to 1 for comparison.

\begin{figure}[t]
    \centering
    \includegraphics[width=0.95\columnwidth]{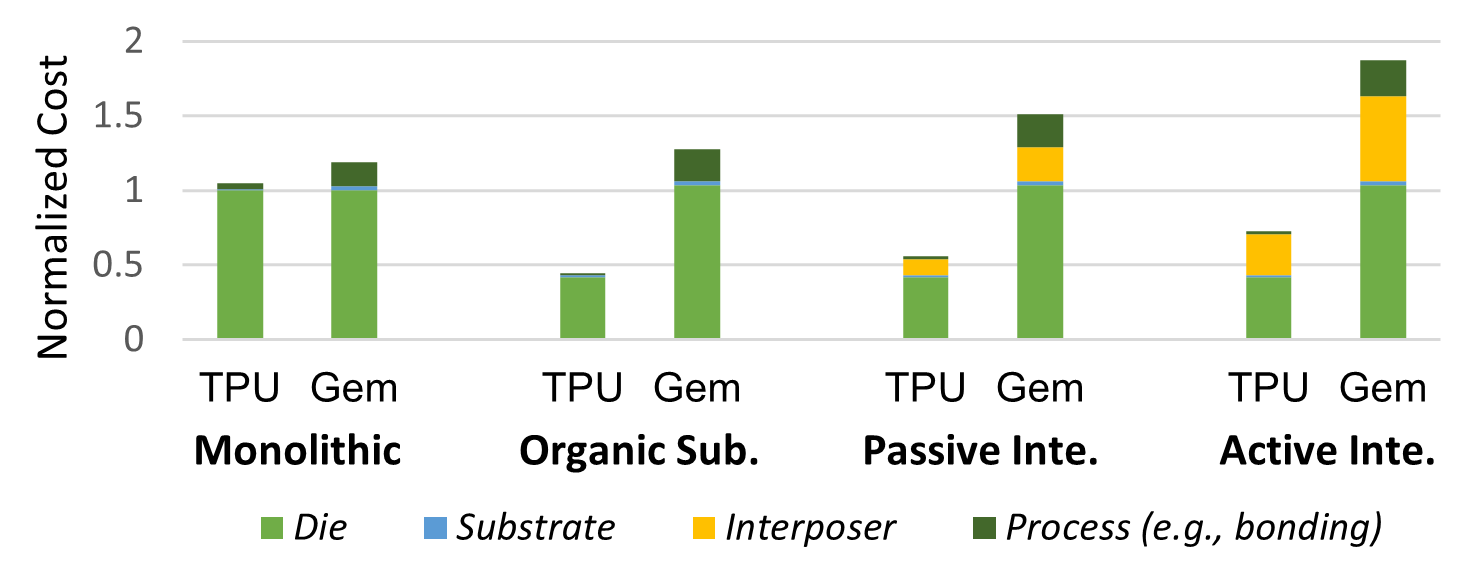}
    \captionsetup{font={small}}
    \caption{Cost comparison for TPU and Gemmini (Gem)}
    \label{fig:costs-compare}
    \vspace{-0.1in}
\end{figure}

From Fig.~\ref{fig:costs-compare}, we can observe that larger chips such as TPU deliver more cost reduction compared to the monolithic die baseline. A smaller chip has a negligible die cost reduction; instead, it requires additional overheads on bonding multiple chiplets. Moreover, the interposer is costly--more than 15\% of the cost is paid on a passive interposer and more than 30\% on an active interposer--for both TPU and Gemmini.

%% file: model.tex
\section{Architecture Modeling}
In this section, we will formulate the modeling problem and then present our framework and performance model.

\subsection{Problem Formulation}
We target modeling highly-specialized chiplet-based accelerators. These accelerators are designed to execute a graph of tensor workloads, with every workload assigned to a specific chiplet. Multiple workloads can be pipelined across different chiplets to overlap their processing time. Our framework can orchestrate multiple chiplets by modeling the non-uniformity in dataflow, pipelining, and communication. In the following, we will formulate the mapping and the non-uniformity.

The mapping refers to assigning where (a chiplet), when (a sequence), and how (a dataflow) to execute workloads:

\begin{figure*}[t]
    \centering
    \subfloat[Dependency graph]{
        \includegraphics[height=0.16\textheight]{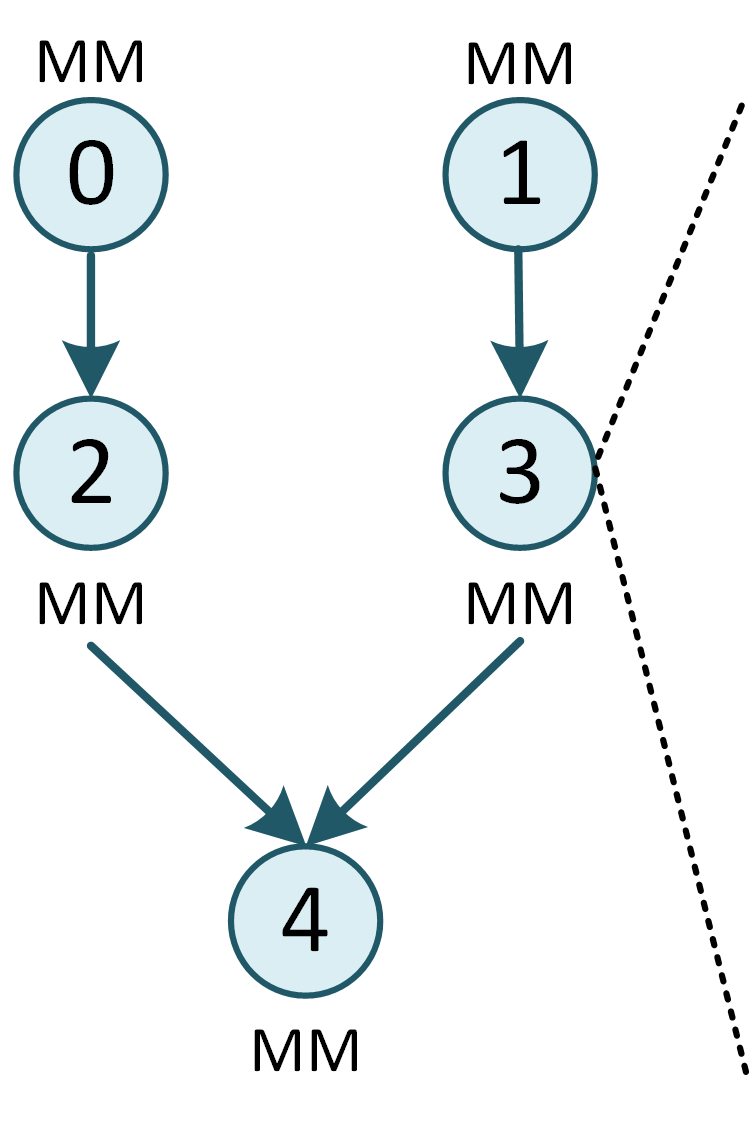}
        \label{fig:workload-dependence-graph}
    }
    \subfloat[Sub-graph of loop instances]{
        \includegraphics[height=0.16\textheight]{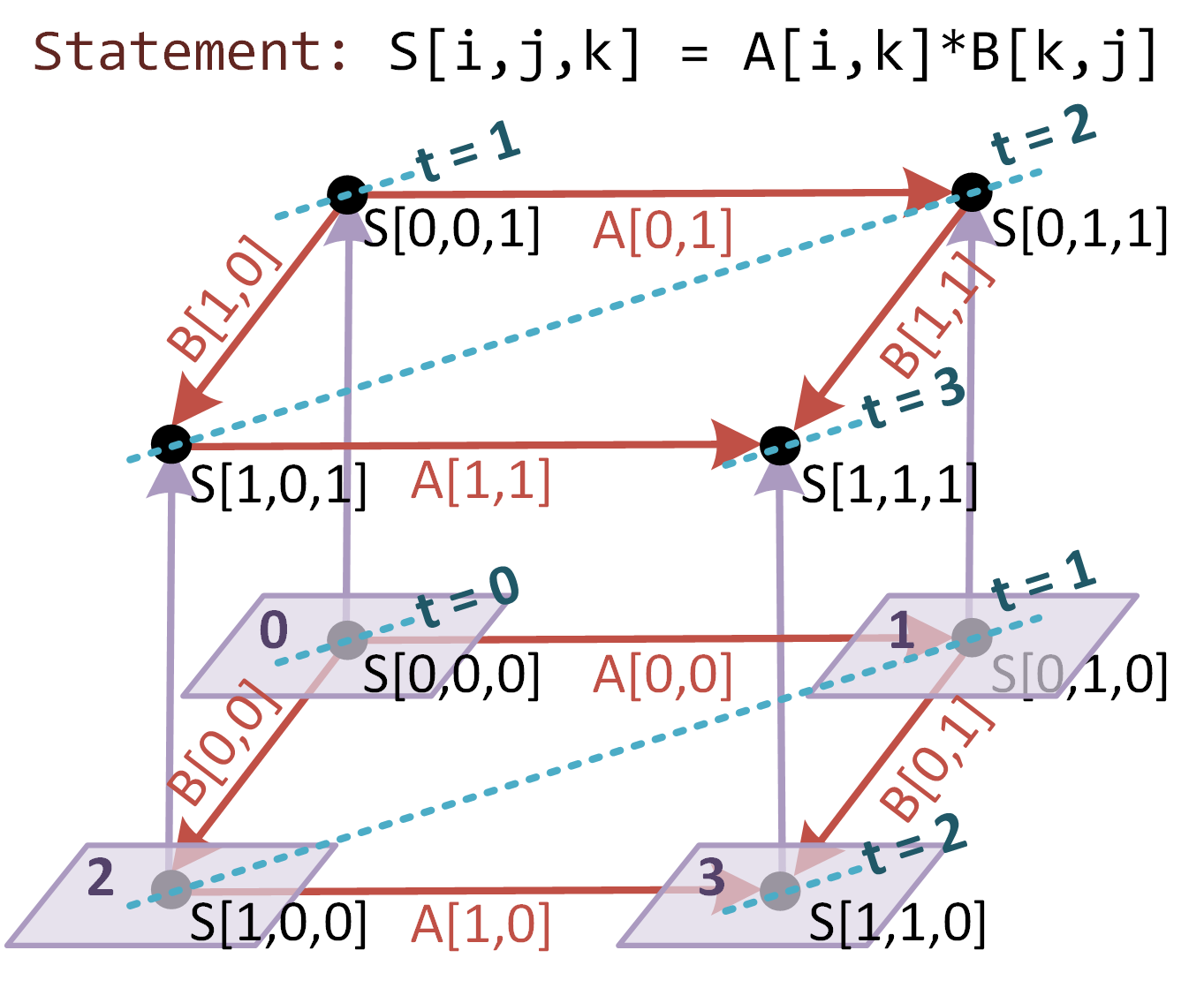}
        \label{fig:fine-grained-syncs}
    }
    \subfloat[Sub-graph of pipelining stages]{
        \includegraphics[height=0.16\textheight]{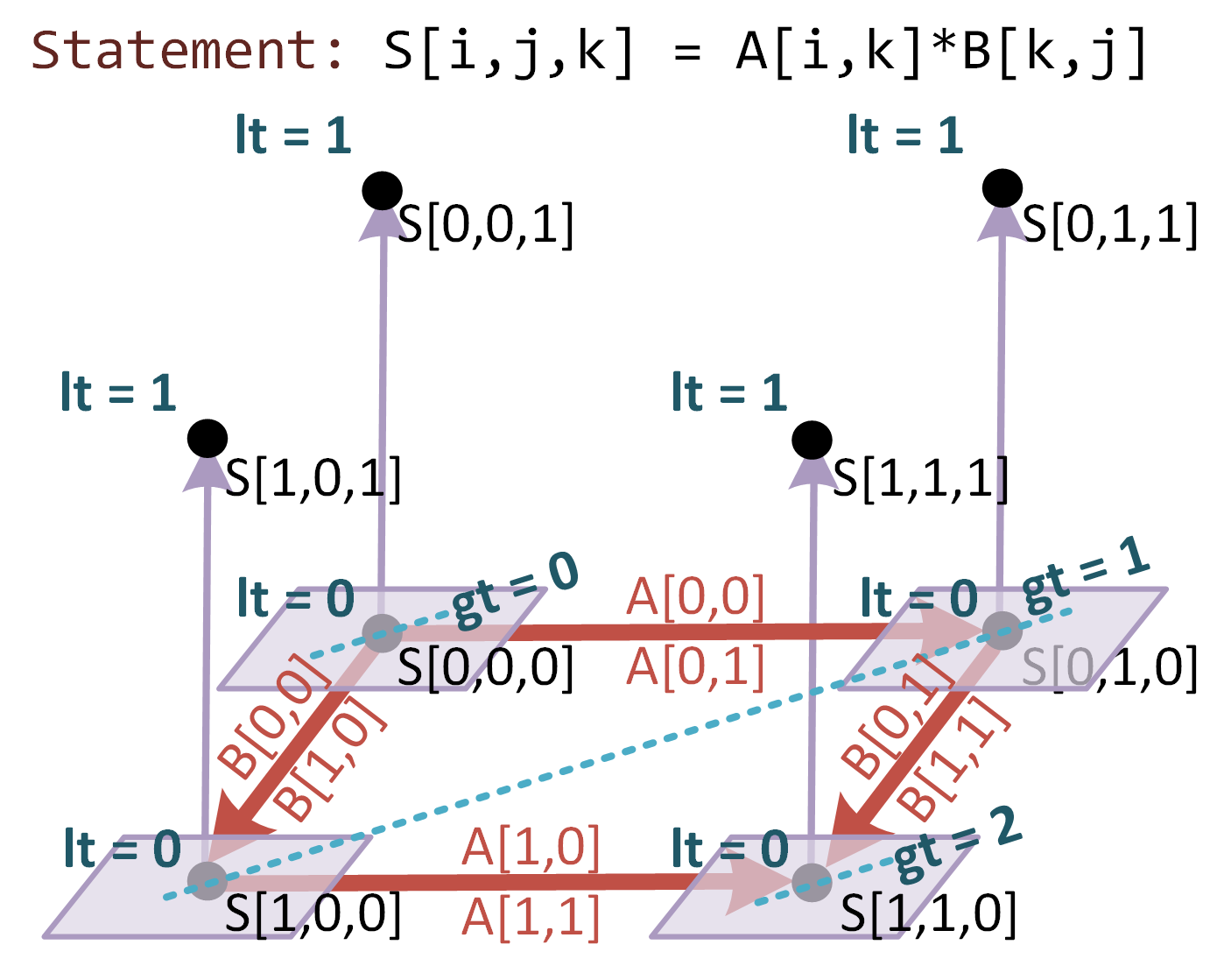}
        \label{fig:coarse-grained-syncs}
    }
    \subfloat[A mapping strategy]{
        \includegraphics[height=0.16\textheight]{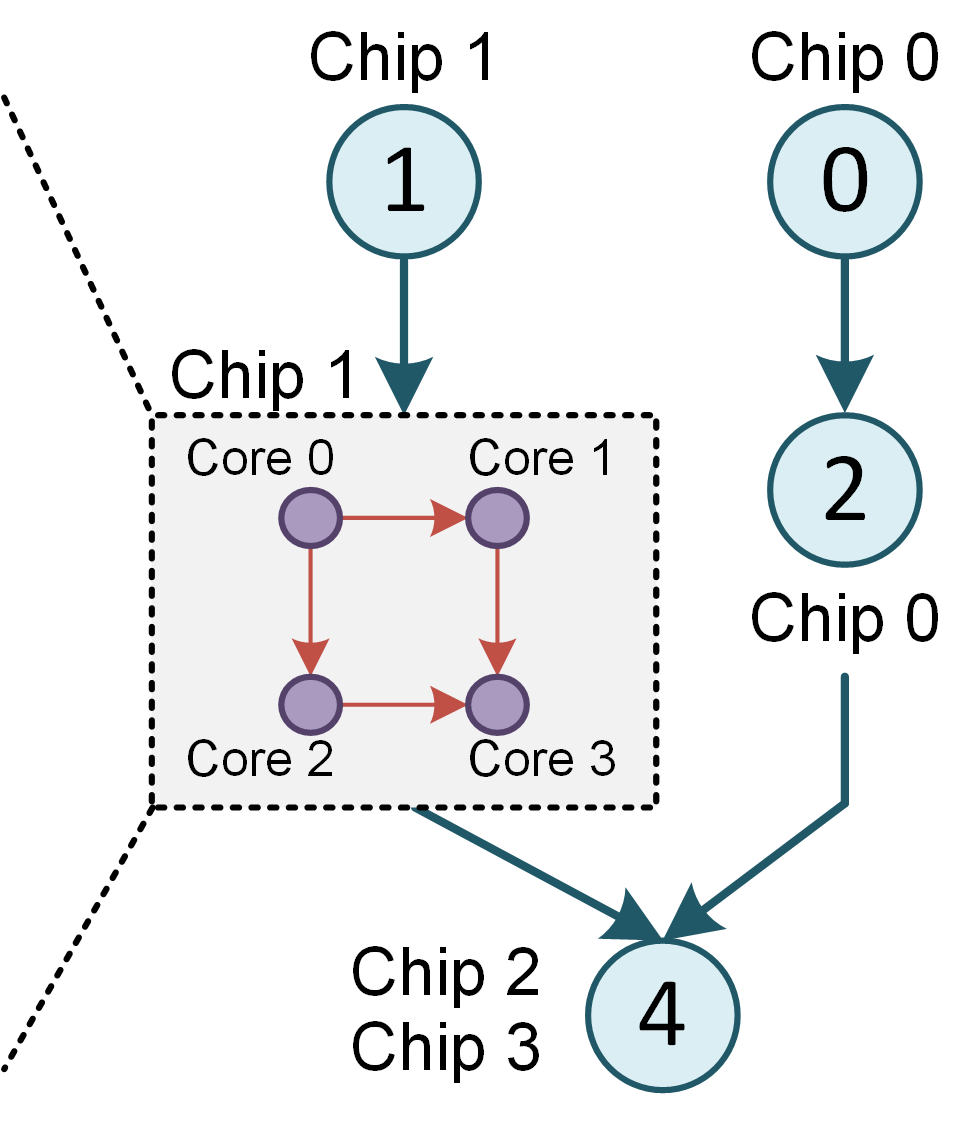}
        \label{fig:communication-graph}
    }
    \captionsetup{font={small}}
    \caption{The mapping formulation. The red arrows represent data transfers, and the dashed blue lines represent pipelining stages.}
    \vspace{-0.1in}
    \label{fig:sync}
\end{figure*}

\textit{Definition 1: }\textbf{Mapping.} A graph $G=(V,\ E)$ has a set of vertices V whose element $v_i$ is a sub-graph $G'$, and a set of edges E whose element $(v_i, v_j)$ indicates $v_j$ depends on $v_i$. At the lowest level, G consists of only one vertex to denote a loop instance. The mapping refers to assigning a vertex $v_i$ onto a computing engine (PE, core, chip). The mapping has to preserve the dependence between vertices.

Fig.~\ref{fig:workload-dependence-graph} gives a Transformer block as an example, which is composed of 2 heads or 5 matrix multiplies (MMs of vertices 0-4). These MMs are mapped to chiplets as shown in Fig.~\ref{fig:communication-graph}. Fig.~\ref{fig:fine-grained-syncs} presents a sub-graph in which every vertex is a loop instance of the matrix multiply statement. A dataflow assigns the instances to be executed on cores (each with one PE for simplicity). This dataflow parallelizes the output matrix (loop $i$ and $j$); thus, input data (e.g., \texttt{A[0,0]}) can be reused and transferred from core 0 to core 1. Another chiplet may adopt a different dataflow or target a different workload. This \textbf{non-uniform dataflow} requires modeling various dataflow and the communication between diverse chiplets.

In Fig.~\ref{fig:fine-grained-syncs}, the time step \texttt{t=0-3} denotes the pipeline stages where each core (PE) computes a loop instance and transfers the reuse value. However, this approach incurs frequent synchronizations and communications between the asynchronous cores, leading to additional overhead and under-utilization of the network bandwidth. To address this issue, we propose to define a hierarchical dataflow as illustrated in Fig.~\ref{fig:coarse-grained-syncs}, where the instances are processed locally (local time \texttt{lt=0,1}) and then the reuse values (\texttt{A[0,0]} and \texttt{A[0,1]}) are transferred at each global time (\texttt{gt=0}) collectively. This approach introduces a sub-graph level, with every vertex denoting the local instances. Generally, a graph hierarchy represents a hardware hierarchy; a high-level graph consists of vertices assigned for chiplets, and low-level graphs for cores and PEs. Hence, the pipeline structure consists of multiple hierarchical levels, with each level composed of its own set of pipelining stages. This \textbf{non-uniform pipelining} requires representing a hierarchy and modeling the possible pipeline stalls.

Communication in a system can be represented as a graph, with each chiplet serving as a network node. DRAM access is treated as a specialized form of communication that originates or terminates at a memory controller. This graph is created to analyze data transfer efficiency by considering the interconnection and available bandwidth provided with an integration design (in-package network and packaging).

\textit{Definition 2: }\textbf{Communication Graph.} A graph $G_c=(N,L)$ with each vertex $n_i\in N$ being a network node and each edge $l_{i,j}\in L$ being the communication flow from $n_i$ to $n_j$, whose weight $bwr_{i,j}$ denotes its bandwidth requirement.

In-package interconnects provide limited bandwidth compared to on-chip ones, exhibiting higher latency and possible performance degradation. Every communication flow (including DRAM access) may experience varying levels of latency due to differences in routing path and resource contention on shared network links. This \textbf{non-uniform communication} can be captured by a contention-aware network model.

The existing frameworks~\cite{magma, nnbaton, atomic} generally model each chip (core) separately and cannot address the three-level non-uniformity. We propose a hierarchical dataflow representation to handle it. Note we target modeling a single execution of the accelerator; only the final results are written to DRAM.


\subsection{Mapping framework}
\label{sec:our-approach}
We first map each workload onto computing engines (PEs, cores, chips) and then analyze their dependencies. We define the computing engine domain for each workload as a cluster.

\textit{Definition 3:} \textbf{Cluster.} A cluster is a domain of computing engines where the loop instances of a workload are mapped. For example, the domain of chiplets in Fig.~\ref{fig:hardware-model} is $Chip[i, j]: 0\leq i,j \leq 1$. The domain of cores and PEs can be defined similarly.

Each workload has its cluster; no workload is mapped to multiple clusters, and no cluster is shared among workloads. Our mapping framework contains three operations. First, we associate a loop instance with a coordinate inside a cluster:
\begin{equation*}
    Map(G, \chi) = \{ v \rightarrow \chi[\vec{p_0}, \vec{p_1}, \vec{p_2}] \},\enspace v \in G.V 
\end{equation*}
where G is a dependency graph with vertex $v$ being a loop instance. $\chi$ is a cluster, whose element $\vec{p}_{0-2}$ is the coordinate of chiplets, cores, and PEs, respectively. This approach gives the flexibility to describe different parallelization options and dataflows. For example, in the case of matrix multiplication, we can parallelize the output matrix with the mapping operation $S[i,j,k]\rightarrow PE[i\%2,j\%2]$ or parallelize the input matrix with $S[i,j,k]\rightarrow PE[i\%2,k\%2]$ on a $2\times 2$ PE array.

Second, we associate a chiplet's coordinate inside a cluster with a chiplet in a system:
\begin{equation*}
	Bind(\chi, C) = \{ \chi[\vec{p_0}] \rightarrow C[\vec{p_1}] \}
\end{equation*}
where $C$ is the domain of chiplets in a system, so it binds a cluster's chiplet $\vec{p_0}$ to system's chiplet $\vec{p_1}$. The binding order indicates the execution sequence. For example, in Fig.~\ref{fig:communication-graph}, we first bind workload 0 onto chiplet 0 then workload 2 onto the same chiplet so that the two workloads are executed sequentially. A cluster also provides logical chiplet interconnects to analyze the data reuse. For example, we can bind workloads on $Chip[0,0]$ and $Chip[1,1]$ in Fig.~\ref{fig:hardware-model}, and these chiplets can be considered directly connected, ignoring the actual network design to build a communication graph. Then from the graph, we derive and analyze the real communication flows.

Third, we generate a hierarchical graph by reducing some vertices into a new vertex under a specified rule:
\begin{equation*}
	Reduce_{r}(G, G’) = \{ v \rightarrow v’ \},\enspace v \in G.V,\, v’ \in G’.V 
\end{equation*}
It reduces multiple vertices $v$ in a graph $G$ into a new vertex $v'$ to generate a new graph $G'$ under a rule $r$. The rule can be set to gather instances mapped to each core, as shown in Fig.~\ref{fig:coarse-grained-syncs}. We can interleave the \textit{map} and \textit{reduce} operations to enable a hierarchical mapping method, i.e., assign a PE-level dataflow (\textit{map}), gather loop instances (\textit{reduce}), then assign a core-level dataflow (\textit{map}), and so on. Finally, we can bind the dataflow to chiplets where a workload will be executed. The operations are performed for each workload in a dependency graph, which directs how the graph is mapped.

We find that two dependent workloads can be pipelined if they are mapped to different chiplets. To make a synchronized execution of multiple workloads, we apply a \textit{reduce} operation on the chiplet-level graph, with the rule gathering instances of a pipeline stage. For example, a batch of feature maps can be pipelined between chiplets processing different DNN layers. A reduced vertex contains computations for a batch and can be assigned with a dataflow. However, a workload dependency graph cannot present data dependencies. For example, workload 4 is mapped to chiplet 2-3 in Fig.~\ref{fig:communication-graph}, so we cannot find to which chiplet the workload 2 and 3's result should be sent (both chiplets are the possible destinations). We can examine the access pattern of a tensor to build data dependency.

Intuitively, if an element is accessed many times, only the first read needs external data to initialize the operation, and only the last write indicates the operations are completed. We denote the set of vertices in $G$ that access an element $F[\vec{f}]$ as $P_{G, F[\vec{f}]}$, which is sorted in its execution order. A set of edges $\Omega$ represents data dependence between two workloads:
\begin{equation*}
	\Omega_{G_1, G_2} = \{ (\max P_{G_1, F[\vec{f}]},\ \min P_{G_2, F[\vec{f}]} )\ |\ \forall \vec{f} \}
\end{equation*}
where $G_1$ and $G_2$ are a producer and consumer, respectively (the connected vertices in a dependence graph), and $F$ is the dependent tensor. This equation examines each element in $F$, and connects the last instance (i.e., \textit{max}) in $G_1$ and the first instance (\textit{min}) in $G_2$ that access the same element, indicating two communicating chiplets.

\subsection{Performance modeling}
\label{sec:perf-model}
\begin{figure}[t]
    \centering
    \subfloat[Pipelining diagram]{
        \includegraphics[height=0.11\textheight]{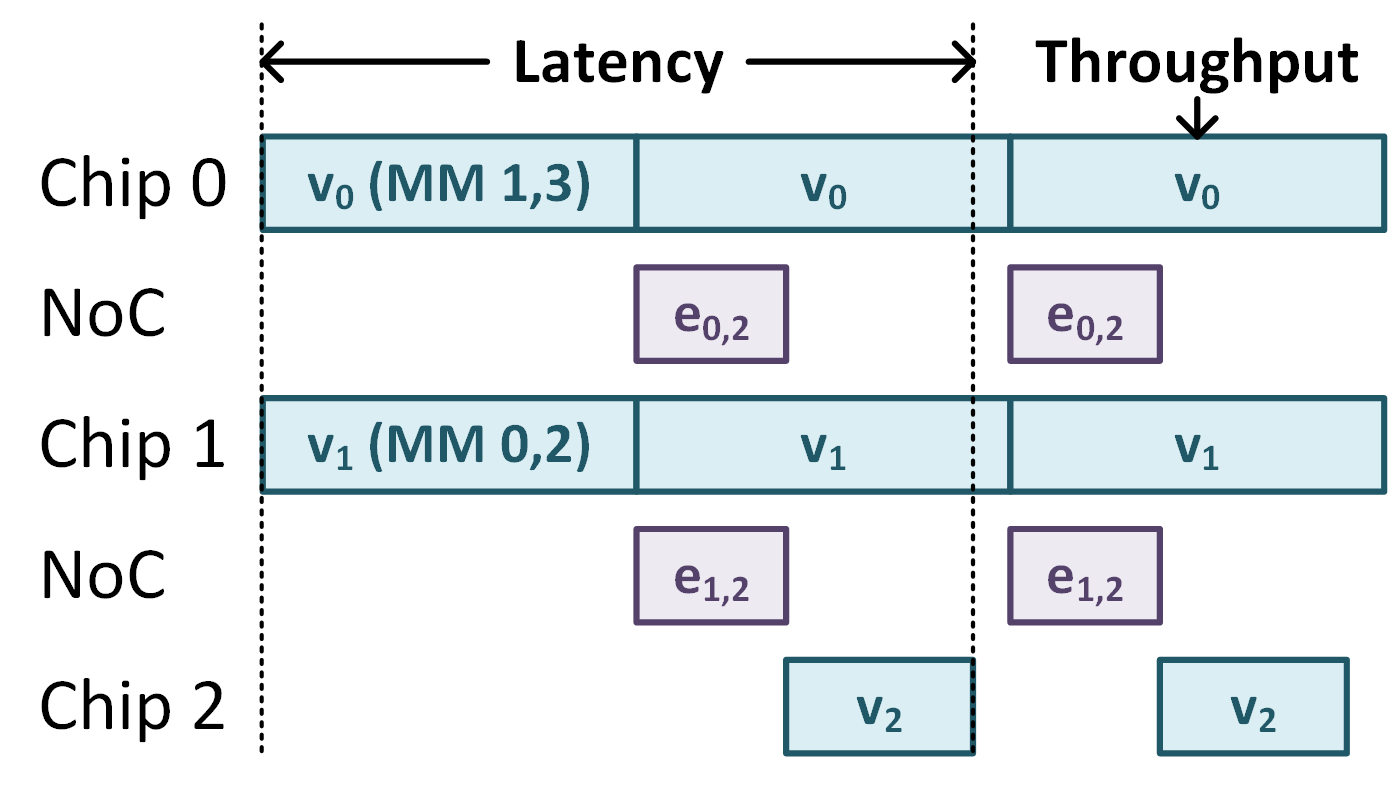}
        \label{fig:pipeline-diagram}
    }
    \subfloat[Data transfer diagram]{
        \includegraphics[height=0.11\textheight]{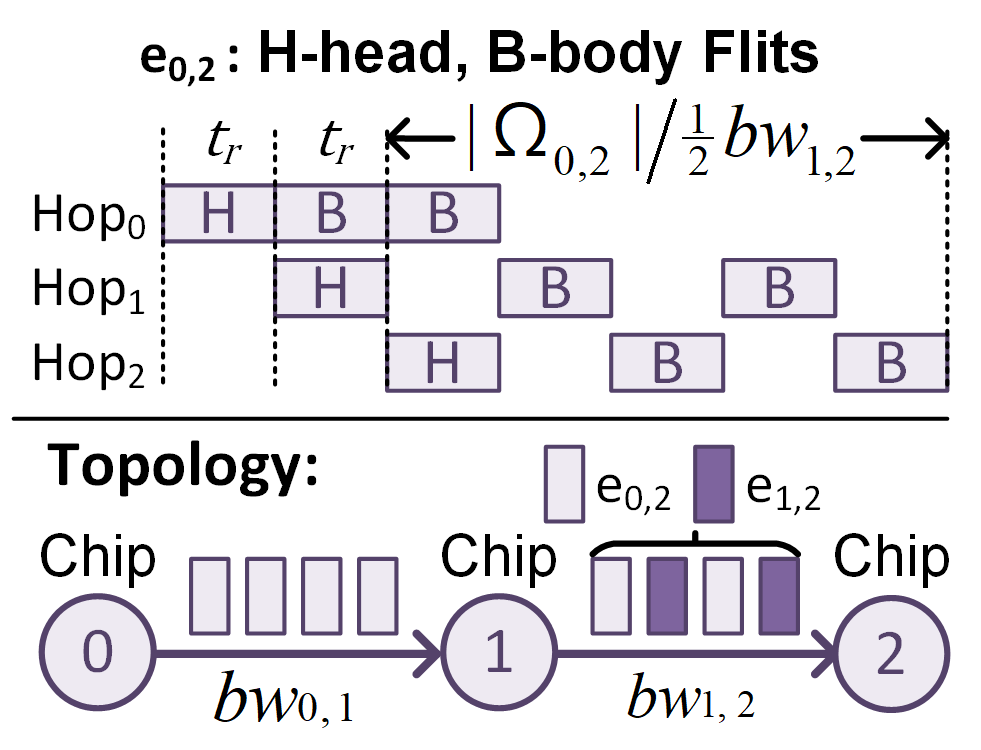}
        \label{fig:noc-diagram}
    }
    \captionsetup{font={footnotesize}}
    \caption{Execution diagrams. $e_{0,2}$ is data transfer from chiplet 0 to 2 and $e_{1,2}$ is data transfer from chiplet 1 to 2. With a linear topology and a fair allocation policy, only half of the link bandwidth can be occupied by $e_{0,2}$.}
    \vspace{-0.1in}
\end{figure}

Our performance model can take a mapping description as input and calculates different metrics by considering pipeline efficiency. The workloads' execution can be pipelined (named as \textit{computing stages}), and data transfer can be pipelined with computing (as \textit{data transfer stages}). The pipeline stages can be derived by traversing the outermost-level graph. For example, we can assume the whole MMs are pipelined in Fig.~\ref{fig:communication-graph}. The workloads mapped to the same chiplet (e.g., MMs 0 and 2) are modeled as a long stage to be pipelined with others. The data transfer stages are added between the two computing stages. The derived pipelining time diagram is shown in Fig.~\ref{fig:pipeline-diagram}.

We can estimate both latency and throughput. The latency refers to the delay to produce an outcome, while the throughput refers to the number of outcomes in a period. Specifically, we can define a path interleaving stages $p=(v_i, e_{i,j}, v_j, ...)$, where $v_i, v_j$ are computing stages and $e_{i,j}$ is a data transfer stage, as shown in Fig.~\ref{fig:pipeline-diagram}:
\begin{equation*}
    \label{equ:metric}
    Lat = \max_{p\in P} \{\sum_{v\in p} D(v) \},\ Thr = \frac{1}{\max_{v\in V}\{D(v)\} }
\end{equation*}
where P is the set of paths and $D(v)$ is the delay of a stage $v$. The latency is the maximal sum of delays on a path, and the throughput is the reciprocal of the longest delay of all stages ($V$ is the set of all stages).

The delay of a computing stage is modeled hierarchically;
\begin{equation*}
    D_C^L = \frac{|V_L|}{U_LN} \times \max\{D_C^{LL_{i}}, D_B^{LL_{i}}, D_A^{LL_{i}}\},\ i\in [0, N]
\end{equation*}
where $L$ is a level (core or chiplet) with $N$ lower-level ($LL$) engines, $|V_L|$ is the number of vertices to be mapped, and $U_L$ is the average utilization of these engines. Every vertex's execution consumes the local processing delay, which is the maximum between computing ($D_C$), memory access ($D_B$), and data transfer ($D_A$) delays. These values can be obtained from a data reuse analysis framework~\cite{tenet,maestro}.

The delay of a data transfer stage $e_{i,j}$ is preferably bounded by the delay of its associated computing stages. We thus derive the bandwidth requirement of it as $bwr_{i,j}=\frac{|\Omega_{G_i,G_j}|}{\min\{D(v_i), D(v_j)\}}$ ($|\Omega_{G_i,G_j}|$ is data transfer volume), and add it into the communication graph. From the graph, our network model estimates delay in given topology, routing, and flow control. The routing algorithm is assumed to be deterministic; it always picks the same path between any two nodes. If multiple flows compete for a link (e.g., $e_{0,2}$ and $e_{1,2}$ compete for the link from chip 1 to 2 in Fig.~\ref{fig:noc-diagram}), a flow control mechanism can throttle data rate to manage resources. We can allocate bandwidth among flows in proportion to their requirements if the sum of them exceeds the available bandwidth. The data transfer delay adds the switch and serialization delay, respectively:
\begin{equation*}
    D(e_{i,j}) = max_{f\in F}\{ |f|\cdot t_s + \frac{|\Omega_{G_i,G_j}|}{min\{ebw_{c}^f\}} \},\ c\in f
\end{equation*}
where $F$ is the set of communication flows associated with $e_{i,j}$. $|f|$ is the hop count and $t_s$ is the delay through one router. As shown in Fig.~\ref{fig:noc-diagram}, the head flit suffers such a delay, but the body flits are pipelined. $ebw_{c}^f$ is channel $c$'s effective bandwidth for flow $f$ (e.g., $ebw_{1,2}^{f}=\frac{1}{2}bw_{1,2}$ in Fig.~\ref{fig:noc-diagram}), and the minimum throttles a flow.

%% file: optimization.tex
\section{Optimization Framework}
The proposed co-optimization framework can encode architecture and integration design parameters, and optimize both with an ML-based exploration engine.

\begin{figure*}[htbp]
    \centering
    \subfloat[Optimization flow with two stages and our encoding schemes] {
        \includegraphics[height=0.2\textheight]{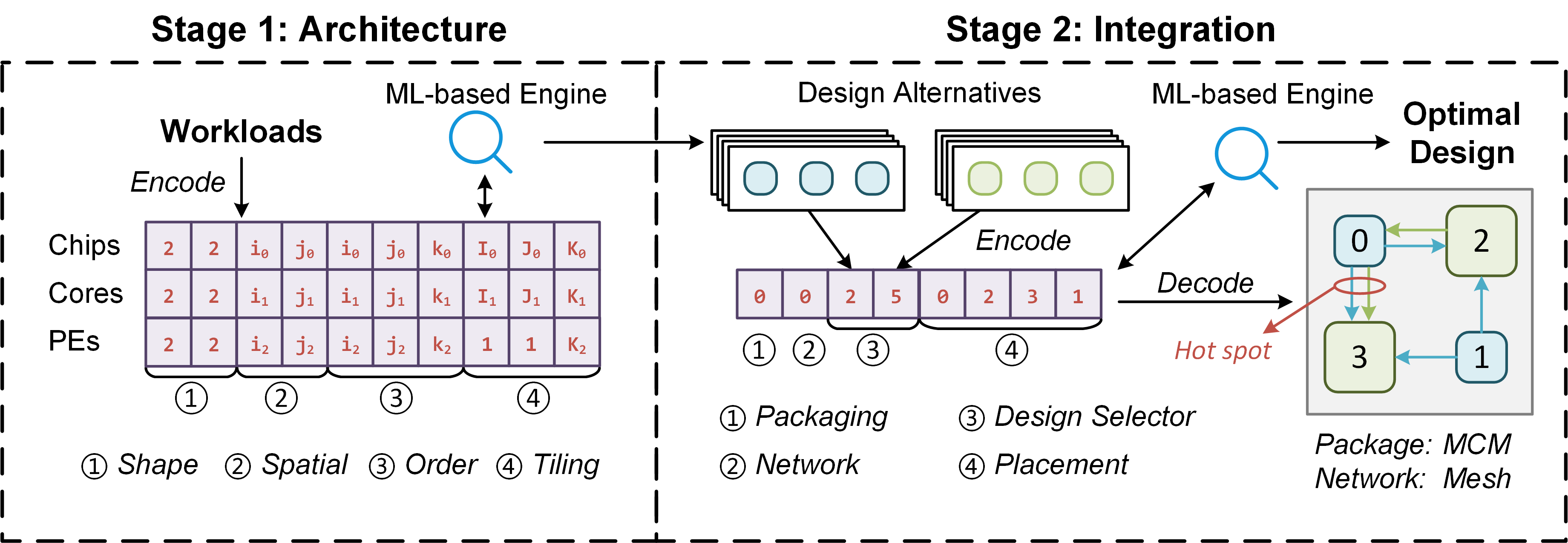}
        \label{fig:opt-flow}
    }
    \subfloat[Bayesian algorithm] {
        \includegraphics[height=0.2\textheight]{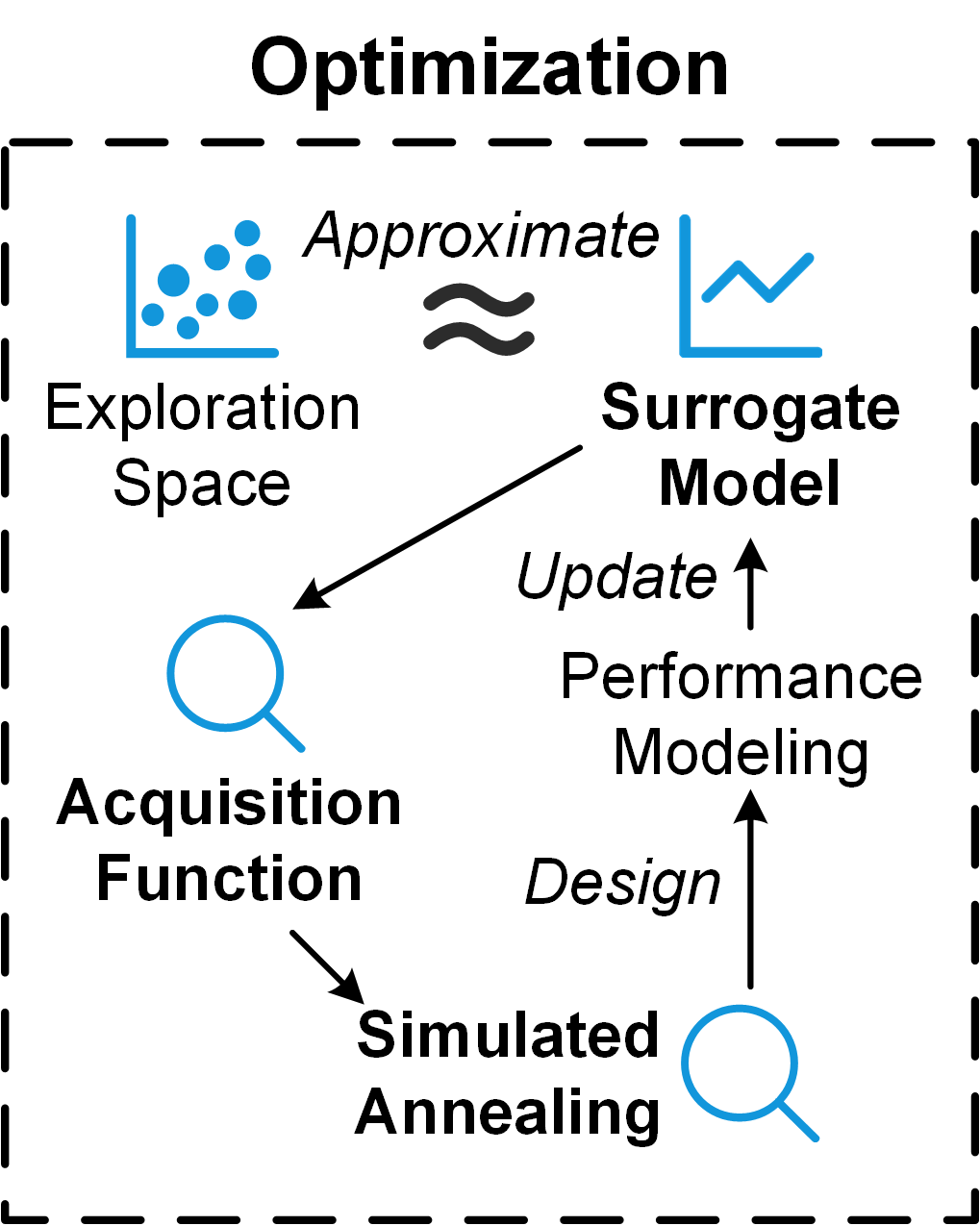}
        \label{fig:opt-engine}
    }
    \captionsetup{font={small}}
    \caption{The illustration of our architecture and integration co-optimization framework}
    \vspace{-0.1in}
\end{figure*}
\subsection{Co-optimization Flow}
\label{ref:co-opt-flow}
Our framework is designed to explore an accelerator executing multiple tensor workloads in a single run. Optimizing a whole application, like a CNN model, might require partitioning the application into multiple runs with variable input shapes and co-optimizing those runs to maximize the end-to-end performance, which is not the focus of our work. 

The proposed co-optimization flow consists of two stages, i.e., architecture and integration stages, as shown in Fig.~\ref{fig:opt-flow}. The architecture stage explores chiplet designs and keeps the pareto optimal ones. The co-design is enabled by encoding the available designs in the integration stage. We then explore the integration-related design aspects. These stages are equipped with an ML-based optimization engine for high sample efficiency. The optimization can make tradeoffs between performance, energy, cost, and area. It evaluates each sampled point using our models. The energy or area is estimated by adding overheads on MACs, memories, and networks.

\subsection{Encoding Scheme}
\label{sec:encoding}
The architectural optimizer accepts a workload description (the dependency, statement, and tensor shapes) and explores chiplet designs where the given workloads are mapped. The encoding scheme we used is illustrated in Fig.~\ref{fig:opt-flow}:

\begin{itemize}
    \item \textbf{Shape.} The geometry of PE, core, and chiplet array.
    \item \textbf{Spatial.} Loops that are assigned for spatial parallelization. Every iteration of these loops is parallelized across a level of computing engines.
    \item \textbf{Order.} The permutation of loops indicates the execution order of these loops.
    \item \textbf{Tiling.} The tile size, e.g., $[I_0, J_0, K_0]$ for loops $i_0, j_0, k_0$ indicates a tile $A[I_0,K_0],B[K_0, J_0]$ to be processed in every chiplet.
\end{itemize}

The \textit{shape} and \textit{tiling} fields correspond to resource assignment, and the \textit{spatial} and \textit{order} fields correspond to dataflow.  We set the buffer size for each tensor based on the tile size. Each level (chiplets, cores, PEs) has its individual parameters, and can be translated into a mapping operation for modeling. The encoding strategy is designed for a single workload, but multiple workloads will be jointly optimized. We can specify the loop (e.g., a batch), inside of which would be a pipeline stage. The buffer size is altered accordingly to accommodate the intermediate data transferred through interconnection. In addition, the throughput of pipelined workloads is limited by the one with maximum computations. To improve efficiency, we can assume a total number of PEs and assign PEs to each workload (chiplet) based on the amount of computations, so that the time spent on each stage is roughly balanced.



The integration optimizer explores different packaging designs. The encoding scheme we used is illustrated in Fig.~\ref{fig:opt-flow}. 
We assign a unique number to the design alternatives (\textit{design id}) as well as the chiplets in the selected design (\textit{chiplet id}).

\begin{itemize}
    \item \textbf{Packaging.} The available options, organic substrate, passive and active interposer, are encoded as 0-2.
    \item \textbf{Network.} Each network design encodes an identifier that represents topology and routing.
    \item \textbf{Design Selector.} Each workload encodes a \textit{design id}.
    \item \textbf{Placement.} Each network node encodes a \textit{chiplet id}. 
\end{itemize}

In Fig.~\ref{fig:opt-flow}, the organic substrate packaging and mesh network are chosen. The 2nd and 5th designs are chosen (in the \textit{design selector} field) for integration, and chiplets in the two designs are numbered; chiplets 0-1 are from the 2nd design, and 2-3 are from the 5th design. The network nodes are also numbered along columns then rows, and the \textit{placement} field $[0,2,3,1]$ indicates that these chiplets are placed to node 0-3, respectively. Some cases may be skipped in exploration, e.g., the total number of chiplets exceeds that of network nodes.


After decoding the information from both architecture and integration stages, we can analyze the communication pattern in a system (as presented in Fig.~\ref{fig:opt-flow}) and evaluate the system-level performance in terms of energy, cost, and area. We set network bandwidth based on the specific requirement of each communication flow. Specifically, the bandwidth is given by summing up the requirements on every network channel and set as the highest requirement among all the channels (i.e., a hotspot). We then adjust the allocated I/O resources, leading to significant variations in area and cost among designs.

It reserves more area for data I/O in a chiplet to support higher bandwidth. Bandwidth density $D_{bw}$ offers the allowed bandwidth per area ($GB/s/mm^2$), varying with packaging and die-to-die interface~\cite{ucie}. The reserved area is estimated as $\frac{bw}{D_{bw}}\times N_{link}$, in which $bw$ is the bandwidth, and $N_{link}$ is the number of links through bumps. For organic substrate and passive interposer, routers are inside chiplets, so all the links pass through bumps. For active interposer, routers are inside the interposer, so only two links connected to a chiplet pass through bumps. The reserved area cannot be scaled with advanced technology and thus hinders cost reduction.

\subsection{Optimization Engine}

Our encoding strategy exhibits a multi-dimensional design space that poses challenges for optimization. In our uniform encoding, we observe that some fields have fixed, relatively-low dimensionalities while others may have high dimensionalities. In the architecture stage, fields \textit{shape} and \textit{spatial} are always two dimensions, while \textit{order} and \textit{tiling} are dependent on the specific workloads. For example, a convolution has 7 loops and 42 dimensions in total for both fields across three levels. In the integration stage, field \textit{packaging} and \textit{network} are identifiers, \textit{design selector} is small with a few workloads, and \textit{placement} depends on network nodes (up to 36).

Fig.~\ref{fig:opt-engine} illustrates our optimization engine that combines a Bayesian (Bayes) and a simulated annealing (SA) algorithm. Bayes exhibits significant efficiency in optimizing an expensive black-box function. In our design, Bayes samples points for the low-dimensional fields, and the function is evaluated by executing an SA engine to optimize the high-dimensional fields and estimating performance. It finds the global optimal in a few attempts by adding prior information to a surrogate model and picking a point at every step with an acquisition function. We take a \textit{Gaussian Process} as the surrogate model and a \textit{probability of improvement} as the acquisition function, which is widely used for its high efficiency~\cite{hasco}. SA exhibits more randomness and suits for high-dimensional fields.

\REM{
We use standard GA operators to generate offspring individuals. The mutation operator randomly chooses genes and mutates them. The crossover operator randomly chooses sites and exchanges genes between the two individuals at these sites. Each chromosome corresponds to an integration solution; the system is shown in Fig~\ref{fig:opt-flow}, for example, after decoding the chromosome.

We sample a point in the encoded space with an efficient simulated annealing (SA) algorithm. Each point can be decoded as an integration strategy, as shown in Fig.~\ref{fig:opt-flow}, which is fed into the network and packaging design modules. After deriving the data transfer latency, the design is evaluated with our performance model under the pipelined execution.}

%% file: experiments.tex
\section{Evaluation}

\REM{
We first re-emphasize the large space of architectural design and packaging technologies, enabled by the powerful relation-centric representation and hardware cost models. Additionally, we demonstrate that heterogeneous chiplet clusters are a promising way to improve performance. Then we show that co-designing dataflow and packaging can further improve the performance. Finally, we evaluate several large-scale real-world applications as a case study.}

\label{sec:exp-setup}

\subsection{Experimental Setup and Validation}
We utilize TENET~\cite{tenet} to analyze data reuse, and Accelergy~\cite{accelergy} to estimate area and energy. The fabrication cost is from ICKnowledge~\cite{icknowledge} under 28nm. The die-to-die interface is UCIe~\cite{ucie}. We validate the proposed modeling framework against ScaleSim~\cite{scalesim}, a systolic array simulator. We assume a four-chip accelerator running Transformer, where each chip has an 8$\times$8 PE array. The latency estimated by our modeling is within 9.8\% error against the simulation results.


\subsection{Comparison}
We compare Monad with the state-of-the-art chiplet-based DNN accelerators, i.e., Simba~\cite{simba} and NN-Baton~\cite{nnbaton}, by realizing their hardware configurations (the same number of PEs and die-to-die interfaces) and mapping strategies in our framework. We collect workloads from typical DNN models, \textit{res[2-5]b\_branch2b} convolution layers from Resnet-50, and four matrix multiply shapes from BERT-large. The parameters are searched with our optimizer.

\begin{figure}[t]
    \includegraphics[width=\columnwidth]{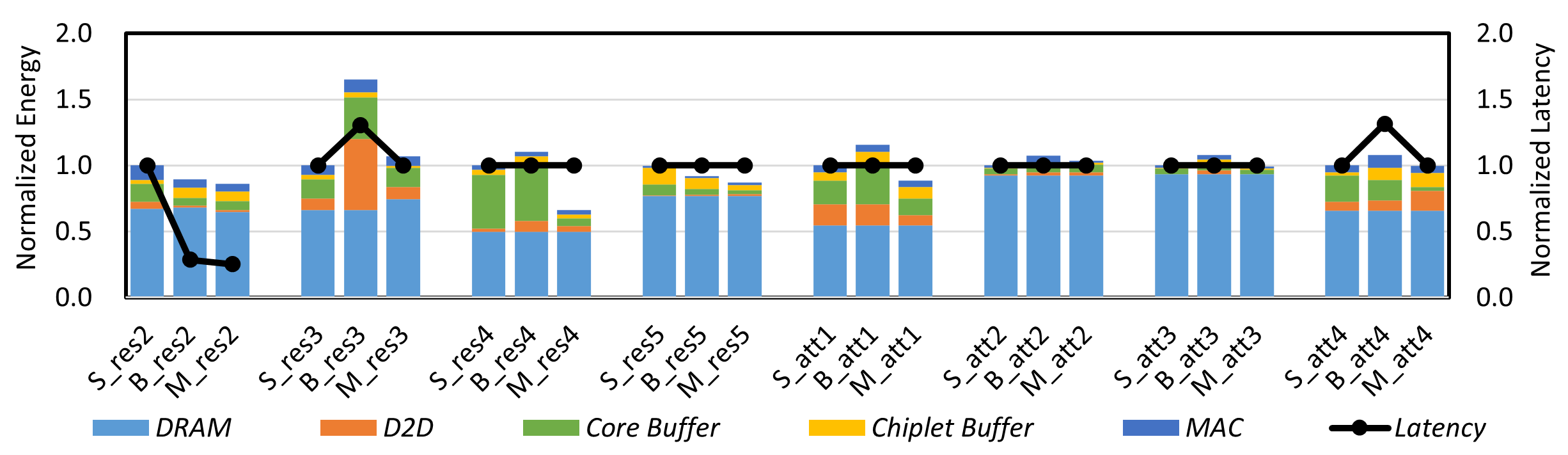}
    \captionsetup{font={footnotesize}}
    \caption{The optimization comparison. We label each group with a prefix (S for Simba~\cite{simba}, B for NN-Baton~\cite{nnbaton}, and M for Monad) followed by the workload: res[2-5] are from ResNet-50, and att[1-4] are from BERT-large.}
    \label{fig:exp1}
    \vspace{-0.1in}
\end{figure}

We adopt energy-delay product (EDP) as our optimization objective. The comparison of energy and latency, along with the breakdown of energy, is shown in Fig.~\ref{fig:exp1}. The results are normalized to that of Simba in every setting. \textit{We achieve an average of $16\%$ and $30\%$ EDP reduction compared to Simba and NN-Baton, respectively.}




We achieve an average of 8\% and 20.8\% energy reduction compared with Simba~\cite{simba} and NN-Baton~\cite{nnbaton}, respectively. NN-Baton uses a ring network where data are rotated among chiplets. It chooses the input with less volume to be reused. For example, in activation-intensive layers (with large feature maps), it reuses weights among chiplets. This strategy gains improvements in settings with unbalanced data volume, such as \textit{res2}, to transfer fewer data and exhibit lower energy than Simba. However, when both inputs are large, as seen in \textit{res3}, the die-to-die data transfer overhead is obtrusive. For latency comparison, Simba maps loop instances by dividing input and output channels. As a result, it may suffer a long latency on the earlier layer (e.g., \textit{res2}) due to inadequate parallelism on channels. On the other hand, NN-Baton chooses a mapping strategy that divides an output plane, making it less effective in cases with inadequate parallelism on outputs.

In summary, each workload has its own optimal microarchitecture design. We can specialize the design for a specific workload to gain improvement, and our modeling framework is general to encompass various design options.

\subsection{Design Space Analysis}
\label{sec:design-space}

\begin{figure}[t]
    \centering
    \subfloat[High performance]{
        \label{fig:latency-opt}
        \includegraphics[height=0.11\textheight]{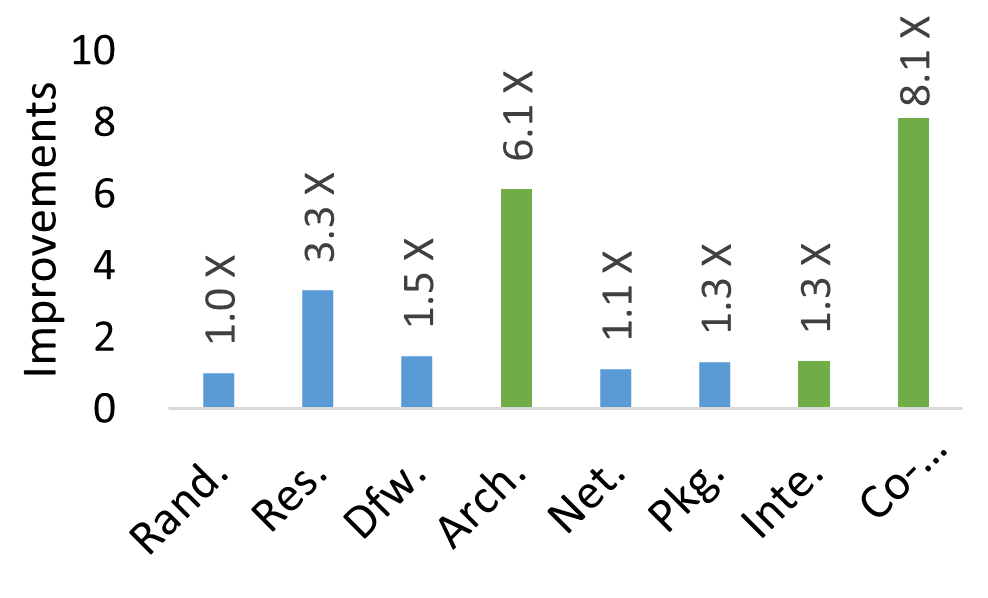}
    }
    \subfloat[Energy efficiency]{
        \label{fig:energy-opt}
        \includegraphics[height=0.11\textheight]{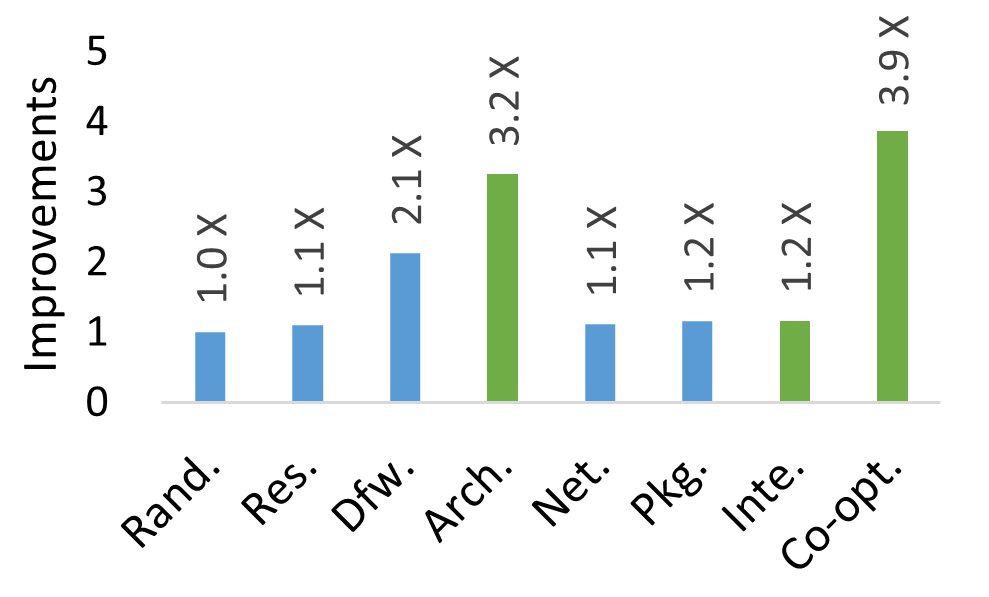}
    }
    \captionsetup{font={footnotesize}}
    \caption{The co-design space analysis. We label each setting as follows: \textit{Rand}-Random, \textit{Res}-Resource, \textit{Dfw}-Dataflow, \textit{Arch}-Architecture, \textit{Net}-Network, \textit{Pkg}-Packaging, \textit{Inte}-Integration, \textit{Co-opt}-Co-optimization.}
    \label{fig:co-design}
\end{figure}

Our proposed co-design approach demonstrates an entangled design space. To illustrate the co-design efficiency, we separately enable different optimizations and their combinations and subsequently report the improvements achieved by each specific setting. Specifically, we evaluate a Transformer block by mapping each matrix multiply workload to chiplets and exploring designs optimized for performance and energy efficiency, respectively. To establish the baseline, we assume a \textit{random} setting using a Simba-like hardware configuration with random parameters. The relative improvements achieved by enabling different optimizations are reported in Fig.~\ref{fig:co-design}. \textit{We achieve 24\% less latency or 16\% less energy compared to the best of separate architecture or integration optimization.}

\begin{figure}[t]
    \includegraphics[width=\columnwidth]{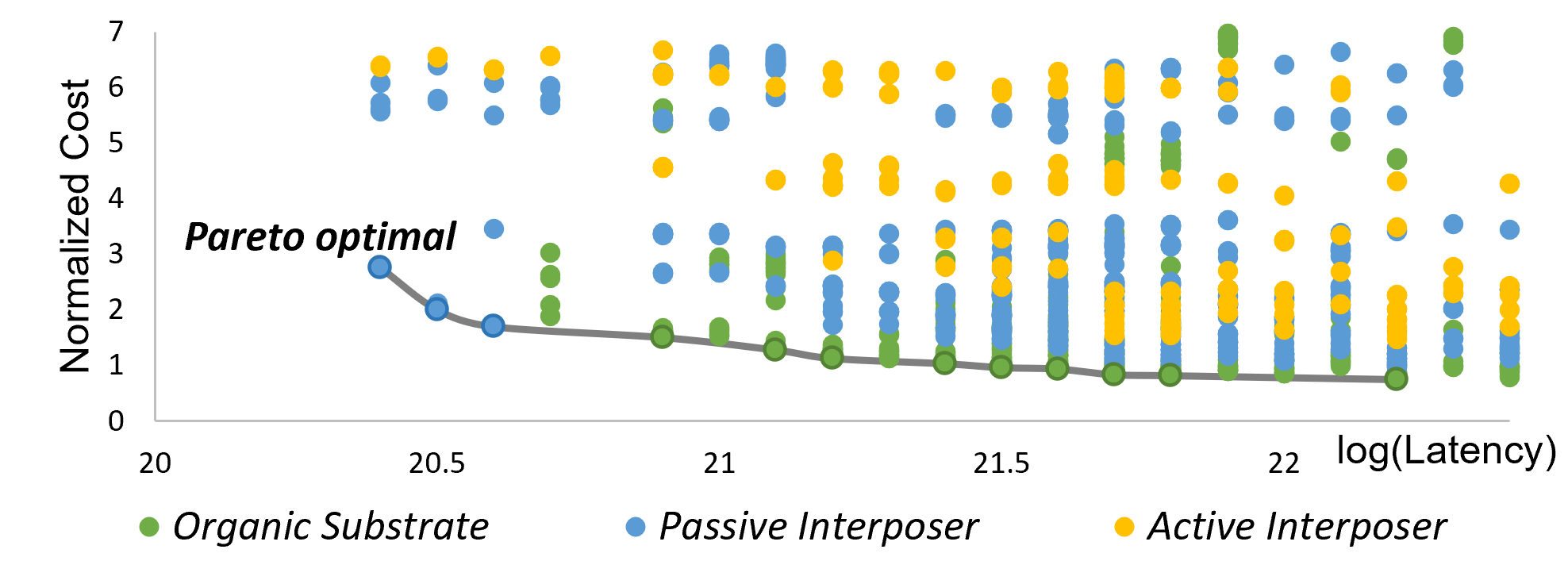}
    \captionsetup{font={small}}
    \caption{Sampled design points and a Pareto optimal analysis for the tradeoffs between cost and latency.}
    \label{fig:pareto}
    \vspace{-0.1in}
\end{figure}

We consider \textit{end-to-end latency} as the performance metric. As shown in Fig.~\ref{fig:latency-opt}, the architecture optimization (\textit{4th} bar), realized by combining resource assignment and dataflow, can result in $6.1\times$ latency reduction by exploiting more PEs and improved utilization. The integration optimization (\textit{7th} bar), achieved by combining network and packaging, leads to $1.3\times$ latency reduction. Our bandwidth setting varies depending on the specific hotspot, and as discussed in $\S$~\ref{sec:perf-model}, data transfer delays contribute to the overall latency. Varying network and packaging can deliver a new hotspot and affect data transfer delays. Overall, co-optimizing computing and communication shows a remarkable $8.1\times$ latency reduction (\textit{8th} bar). 

The results for energy efficiency are illustrated in Fig.~\ref{fig:energy-opt}. The architecture optimization achieves $3.2\times$ energy reduction by exploring tradeoffs between two interrelated aspects; allocating a larger buffer increases energy consumption, but also facilitates dataflow exploration to gain benefits from on-chip data reuse. The integration optimization leads to $1.2\times$ energy reduction by reducing the number of hops in data forwarding, thereby minimizing energy consumption associated with die-to-die interfaces. This can be achieved by changing network topology and placement (\textit{5th} bar), and by varying the chiplets to be integrated (\textit{6th bar}). Overall, the co-optimization leads to $3.9\times$ energy reduction by enabling more tradeoffs.

A cost-effective design approach aims at minimizing cost while keeping the same level of performance. We still adopt a Transformer block but optimize for cost-effectiveness. Some sampled design points classified in their packaging technologies are shown in Fig.~\ref{fig:pareto}. There is a wide range of costs (up to $7\times$) in the same level of latency, which indicates varying levels of resource utilization and bandwidth requirements. A costly interposer can reduce the overall cost by allocating less area for I/O. Therefore, cost optimization necessitates striking a balance between computing and communication efficiency, entailing a comprehensive co-design flow.


\subsection{Case study}
Cost is a crucial objective when exploring the design space of chiplet-based systems. Ignoring cost often leads to a large monolithic design and limited exploration of communication efficiency in advanced packaging. We illustrate these design considerations through a concrete example.

\begin{figure}[t]
    \subfloat[A tensor network for TT]{
        \label{fig:tensor-network}
        \includegraphics[height=0.1\textheight]{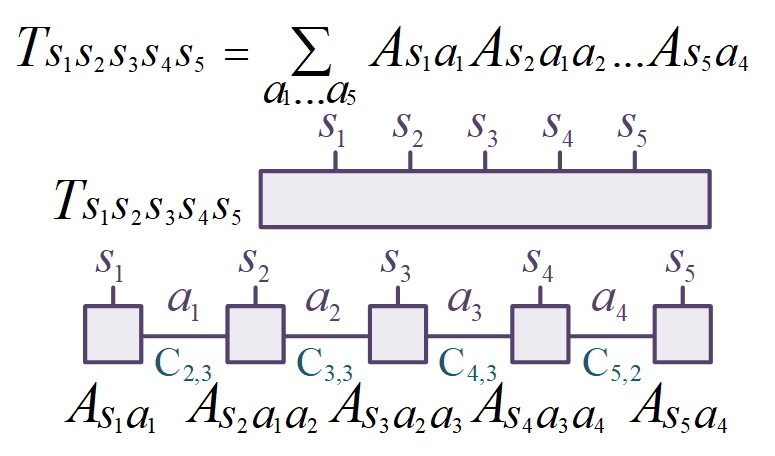}
    }
    \hspace{-0.1in}
    \subfloat[A chiplet-based accelerator]{
        \label{fig:case-design}
        \includegraphics[height=0.1\textheight]{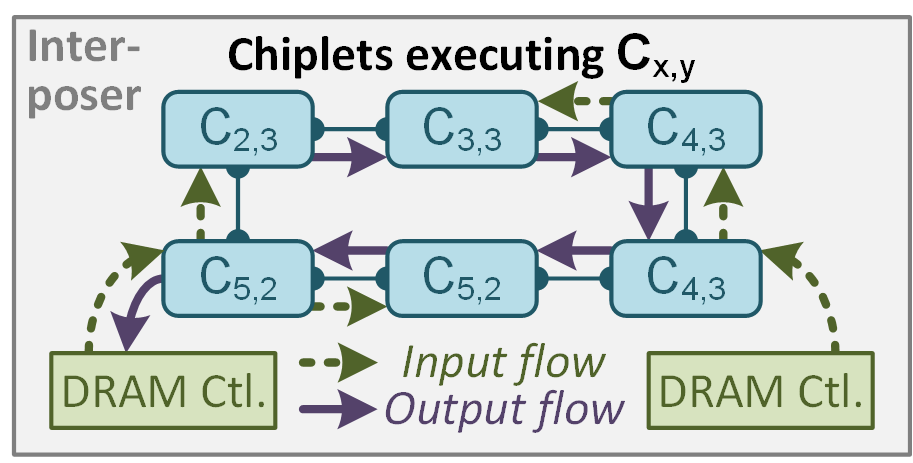}
    }
    \captionsetup{font={footnotesize}}
    \caption{Case study of tensor-train (TT) decomposition. It is represented as a tensor network, where tensors are denoted as rectangles and tensor indices are denoted as lines ($s_{1-5}$, $a_{1-4}$). The line connecting two tensors implies tensor contraction  $C_{x,y}$ ($x,y$ are the number of indices of the two tensors, respectively) over the connected indices ($a_{1-4}$)~\cite{tensor-network}. After that, we will get a new tensor with $x+y-2$ dimensions.}
    \label{fig:case-study}
\end{figure}

Tensor train (TT) decomposition provides a compact representation for high-dimensional tensors and finds applications in  quantum physics and machine learning~\cite{sidiropoulos2017tensor}. As shown in Fig.~\ref{fig:tensor-network}, a 5-dimensional tensor $T$ can be decomposed into a sequence of smaller tensors $A$. The first and last ones, $A_{s_1a_1}$ and $A_{s_5a_4}$, are matrices, and the others are 3D tensors. These tensors are contracted by performing inner products over the shared indices, i.e., a generalized matrix multiply for tensors. For example, we can get a new tensor $A_{s_1s_2a_2}$ by contracting $A_{s_1a_1}$ and $A_{s_2a_1a_2}$ over $a_1$. After one-by-one contraction, the result tensor gradually expands in dimensions until 5.

Fig.~\ref{fig:case-design} presents our accelerator design, which adopts one chiplet for the lower-dimensional workloads ($C_{2,3}$ and $C_{3,3}$) and two chiplets for the higher-dimensional workloads ($C_{4,3}$ and $C_{5,2}$ with $O(n^6)$ computations). Using two chiplets can bring 28\% cost reduction compared to the monolithic design. We observe that when the cost is not taken into account, our optimizer may scale up a chip to avoid additional overheads. Though we can set an area constraint based on potential cost reduction, directly using cost as an objective would consider the fabrication node, yield, and intricate interplay with other components. For instance, we still employ a single chip for $C_{3,3}$ despite its larger size ($>300mm^2$). Solely focusing on area cannot make optimal cost reduction strategies.

To match the communication requirement of the sequential tensor contraction workloads, a ring network is chosen with a passive interposer. Trading cost and performance is critical in determining whether the enhanced connectivity can amortize the cost of advanced packaging. Besides, we observe that the energy consumption decreases with advances in packaging; it becomes lower than that of SRAM (0.25 pJ/bit~\cite{ucie} vs. 0.81 pJ/bit~\cite{nnbaton}). Thus, solely focusing on energy may result in a design that excessively relies on network data transfers with impractical bandwidth requirements. Considering cost makes a direct influence on communication efficiency. For example, the optimized design in Fig.~\ref{fig:case-design} assigns a dedicated channel for each flow to avoid extra costs on I/O resources.

%% file: related-work.tex
\section{Related Work}
\textbf{Multi-chip accelerators.} There has been significant work focusing on the tiled multi-core architecture. TANGRAM~\cite{tangram} proposes dataflow optimization for intra-layer parallelism and inter-layer pipelining. Atomic~\cite{atomic} proposes an optimization framework to map and schedule a DNN model on a scalable accelerator. Simba~\cite{simba} and NN-Baton~\cite{nnbaton} introduce MCM-based DNN accelerators and workload mapping methods for them. SPRINT~\cite{sprint} develops a chiplet-based accelerator with photonic interconnects. Some work instead targets specialized cores. Herald~\cite{herald} and MAGMA~\cite{magma} schedule DNN layers onto the cores with different dataflows for efficiency. Krishnan et al~\cite{krishnan2022big} exhibits an in-memory computing architecture with big-little chiplets to improve utilization. H2H~\cite{zhang2022h2h} and COMB~\cite{comb} propose communication-aware mapping frameworks for heterogeneous systems. However, they commonly consider each chip would run independently and thus model each chip separately, with less focus on their coordination.

\textbf{Chiplet.} A large body of work focuses on interposer-based network designs~\cite{yin-noc,li2022gia,wang-noc,kite}. Several works~\cite{kim-codesign, kabir-codesign} introduce EDA flows that couple architecture and packaging design. Coskun et al.~\cite{coskun-codesign} jointly explore the network topology and chiplet placement. Pal et al.~\cite{dse-chiplet} select suitable chiplets from many candidates to build an efficient processor, but they miss the integration design space. Feng et al.~\cite{feng2022chiplet} and Tang et al.~\cite{tang2022cost} suggest to explore the chiplet granularity, heterogeneity, and reuse strategies using a cost model. In contrast, this paper makes a tradeoff between PPA and fabrication cost by introducing a performance model.

\section{Conclusion}
This paper proposed Monad, an exploration framework for chiplet-based spatial accelerators. We developed an architecture modeling framework to evaluate performance considering the non-uniformities of specialized chiplets. We also proposed to couple the architectural and integration design spaces with an ML-based optimization approach. The experiment demonstrated a significant EDP reduction and a huge tradeoff space by incorporating the cost as a design objective.

\section*{Acknowledgment}
This work is supported in part by the National Natural Science Foundation of China (NSFC) under grant No.U21B2017.